\documentclass{article} 
\usepackage{arxiv}

\usepackage{hyperref}
\usepackage{url}

\usepackage{amsfonts}
\usepackage{amsmath}
\usepackage{float}
\usepackage{bm}

\usepackage{amsthm}

\usepackage[ruled]{algorithm2e} 

\SetAlFnt{\small}
\SetAlCapFnt{\small}
\SetAlCapNameFnt{\small}
\SetAlCapHSkip{0pt}
\IncMargin{-\parindent}
\SetKwInput{KwParam}{Parameters}
\SetKwInput{KwInit}{Initialize}
\DontPrintSemicolon

\usepackage{tcolorbox}
\usepackage[para]{footmisc}

\usepackage{pgfplots}

\newcommand{\N}{{\mathbb N}}
\newcommand{\R}{{\mathbb R}}

\newcommand{\Id}{\ensuremath{\mathbf I}}

\newcommand{\Domain}{\mathcal{D}}
\newcommand{\Range}{\mathcal{R}}
\newcommand{\NDist}{{\mathcal N}}
\newcommand{\UDist}{{\mathcal U}(0, 1)}
\newcommand{\Mech}{{\mathcal M}}
\newcommand{\Bench}{{\mathcal B}}
\newcommand{\E}{{\bf E}}

\newcommand{\eps}{\varepsilon}
\newcommand{\eqdef}{\stackrel{\Delta}{=}}

\newcommand{\eqr}[1]{Eq.~(\ref{#1})}
\newcommand{\algr}[1]{Alg.~(\ref{#1})}

\newcommand{\figr}[1]{Fig.~\ref{#1}}
\newcommand{\tabr}[1]{Tab.~\ref{#1}}

\newcommand{\citep}[1]{(\cite{#1})}

\title{Towards Prior-Free Approximately Truthful One-Shot Auction Learning via Differential Privacy\thanks{We want to thank Olivia Röhrig for her invaluable contributions to increasing the clarity of this document, the reviewers of the DPML workshop at ICLR 2021 for their instructive feedback, as well as NLnet for its manifold organizational support. Code for the experiments is available at: \url{https://github.com/degregat/one-shot-approx-auctions/}}}

\author{
  Daniel Reusche\\
  \texttt{research@degregat.net}\\
  \And
  Nicolás Della Penna\\
  \texttt{nikete@mit.edu}\\
}

\begin{document}

\maketitle

\begin{abstract}

Designing truthful, revenue maximizing auctions is a core problem of auction design. Multi-item settings have long been elusive. Recent work of \cite{DOA} introduces effective deep learning techniques to find such auctions for the prior-dependent setting, in which distributions about bidder preferences are known. One remaining problem is to obtain priors in a way that excludes the possibility of manipulating the resulting auctions. Using techniques from differential privacy for the construction of approximately truthful mechanisms, we modify the RegretNet approach to be applicable to the prior-free setting. In this more general setting, no distributional information is assumed, but we trade this property for worse performance. We present preliminary empirical results and qualitative analysis for this work in progress.

\end{abstract}


\section{Introduction}
Auction design is a core problem of economic theory. The resulting auctions find practical applications for example in spectrum auctions for allocation of frequency bands to wireless carriers, commodity auctions, as well as online auctions platforms like eBay.

In the standard model of {\it independent private valuations}, bidders have valuations over items and utility dependent on the items allocated to them. The auctioneer does not know the realized valuations of the bidders, but has access to aggregate information in the form of distributions over valuations. Since valuations are private, incentivizing bidders to report them truthfully is important for finding revenue maximizing auctions.

\cite{OAD} presents an optimal (truthful and revenue maximizing) auction for the single-item multi-bidder setting, but the multi-item setting has long been elusive for reasons of computational intractability; For a survey on intractability, see the introduction of \cite{ALG}. The last 10 years have seen advances in partial characterizations, algorithmic results, albeit satisfying weaker notions than truthfulness, as well as applications of tools from machine learning and computational learning theory; A survey on these developments can be found in the introduction of \cite{DOA}.

Recently a line of work by \cite{DOA} introduces deep learning techniques to find revenue maximizing, truthful (or: dominant strategy incentive compatible) auctions. It develops the RegretNet approach, where regret, a measure for incentive compatibility, is used to constrain the learning problem of finding revenue maximizing auctions by way of the augmented Lagrangian method. It is able to recover known solutions and finds low-regret solutions for multi-item settings. This approach is limited to the prior-dependent setting though, in which knowledge about the distributions of valuations is assumed. These distributions need to be aggregated in an incentive compatible manner, to prevent influence on the outcome of the mechanism by this avenue. Most recently \cite{ALG} builds on RegretNet, but increases efficiency and applies reductions to get truthful auctions from low regret ones.

There also exist interesting connections between robust mechanism design and differential privacy: \cite{MDvDP} introduce its use as a solution technique for mechanism design, by using the guarantees it provides to bound the influence agents can have on the outcome of mechanisms. These mechanisms are approximately truthful (and approximately optimal), since the bounded influence results in bounded incentives to misreport. \cite{AOMDvDP} further analyze the potential for optimality approximation of mechanisms utilizing differential privacy.

\subsection{Contributions}
To work towards an approach for solving the more general prior-free setting, in which no knowledge about valuation distributions is assumed, we integrate the above techniques.

By using the work of \cite{DLDP}, we make the training of RegretNet differentially private, and thus robust to changes in distribution. This robustness allows us to remove the distribution requirement and enables us to use RegretNet on single bids profiles to perform one-shot learning, giving us one auction per bid profile.

Preliminary empirical analysis leads us to believe that the resulting auctions are approximately truthful, prior-free approximations of optimal mechanisms.

We present a qualitative analysis of computational experiments with the modified codebase and formulate theoretical problems, the solution of which we deem necessary for a thorough characterization.

\section{Background}

As we have seen, it is possible to learn multi-item auctions in the prior-dependent setting. What follows is a more in depth explanation of the techniques we are going to use to tackle the prior-free setting.

\subsection{Auction Design}
The {\bf multi-item auctions} we will consider consist of sets $N = \{1, \dots, n\}$ and $M = \{1, \dots, m\}$ of $n$ bidders and $m$ items. Each bidder $i$ has valuations $v_i(\{j\})$ for all items $j$.
We focus on bidders with {\bf additive valuations}\footnote{For a more general exposition see Section 2 of \cite{DOA}. Section 3 of \cite{ALG} shows another description of the additive setting.}, where valuations for sets of items $S \subseteq M$ are $v_i(S) = \sum_{j \in S} v_i(\{j\})$.

In the {\bf prior-dependent setting}, the auctioneer does not know the valuation profile $v = (v_{1}, \dots, v_{n})$ of the agents in advance, but has prior knowledge about the distribution $F$ from which $v$ is drawn. $V = \R^{n \times m}$ is the space of possible valuations $v$ and bids $b$, with $v_i, b_i \in V_i$ and $V_i = \R^m$. Bidders report their bids, $b = (b_1, \dots, b_n)$ with $b_i = v_i$ being a {\bf truthful report} and $b_i \neq v_i$ being a {\bf misreport} of agent $i$.

Each auction is defined by a pair of allocation and payment rules $(g, p)$ with $g_i : V \to [0, 1]^m$ giving allocation probability of each item to agent $i$ and $p_i : V \to \R_{\geq 0}$ giving the necessary payment. Any item is allocated at most once: $\sum_i g_i(b) \leq 1$ for all $b \in V$. Bidder $i$, with valuation $v_i$, receives {\bf utility} $u_i(v_i;b) = v_i(g_i(b)) - p_i(b)$ for a set of bids $b$ from all bidders. The {\bf revenue} of an auction is $\sum_{i \in N} p_i(v)$.

Further, let $v_{-i}$ be the valuation profile $v$ without $v_i$. $b_{-i}$ and $V_{-i}$ are used analogously. An auction is {\bf dominant strategy incentive compatible (DSIC)} if a bidders utility is maximized when reporting truthfully, independent of the bids of others: $v'_i \neq v_i$ being a misreport, $u_i(v_i; (v_i, b_{-i})) \geq  u_i(v_i; (v'_i, b_{-i}))$ holds for all $v_i, v'_i \in V$ and bids of others $b_{-i} \in V_i$.

An auction is ex post {\bf individually rational (IR)} if each bidder always receives non-negative utility: $u_i(v_i;(v_i, b_{-i}) \geq 0$ for all $v_i \in V_i, b_{-i} \in V_{-i}$.

Optimal auction design seeks a DSIC auction that maximizes revenue and is IR.

\subsection{Optimal Auction Design with Deep Learning}
To translate optimal auction design into a {\bf learning problem} (Section 2.2.2 of \cite{DOA}) we take a parametrized class of auctions $(g^w, p^w)$, with parameters $w \in \R^d, d \in \N$.

With expected ex post {\bf regret} being the utility gain for optimal misreports, and $u^w_i(v_i; b) = v_i(g^w_i(b)) - p^w_i(b)$, we can measure the deviation from DSIC:
\begin{equation}\label{eq:regret}
  rgt_i(w) = \E[\max_{v'_i \in V_i} u^w_i(v_i;(v'_i, v_{-i})) -  u^w_i(v_i;(v_i, v_{-i}))].
\end{equation}
Thus an auction is DSIC iff $rgt_i(w) = 0, \forall i \in N$, except for measure zero events. We then minimize expected negated revenue, subject to a regret constraint for each bidder and $F$ being the valuation distribution:
\begin{equation}\label{eq:learning}
  \min_{w \in \R^d} ~ \E_{v \sim F}\Big[-\sum_{i\in N}p^w_i(v)\Big]  ~ \text{s.t.} ~ rgt_i(w) = 0 ~ \forall i \in N, v \in V
\end{equation}

For implementation with a deep learning pipeline, we formulate the {\bf empirical regret} for a sample of L valuation profiles as
\begin{equation}\label{eq:emp-learning}
\widehat{rgt}_i(w) = \frac{1}{L}\sum_{l=1}^L \Big[\max_{v'_i \in V_i} u^w_i\big(v_i^{(l)}; \big(v'_i, v^{(l)}_{-i}\big)\big) - u^w_i(v_i^{(l)}; v^{(l)})\Big]
\end{equation}
as well as the empirical loss, which we want to minimize, subject to an empirical regret constraint:
\begin{equation}\label{eq:emp-loss}
\min_{w \in \R^d} ~ -\frac{1}{L}\sum_{l=1}^L \sum_{i=1}^n p^w_i(v^{(l)}) ~ \text{s.t.} ~ \widehat{rgt}_i(w) = 0 ~ \forall i \in N, v \in V
\end{equation}

The allocation and payment rules are modeled as neural networks (Section 3.2 of \cite{DOA}), which ensure IR by restricting to auctions which don't charge any bidder more than their valuations for any allocation. This implementation is called RegretNet.

The regret constraint can be incorporated into the objective by using the technique of Lagrange multipliers (Section 4 of \cite{DOA}).

\subsection{Prior-Free Approximations to Optimal Mechanisms}

A {\bf prior-free mechanism} (Section 7 of \cite{MDnA}) is a mechanism (Section 9.4 of \cite{AGT}) which does not make assumptions about the valuation distribution of agents. In the case of auction design, this would be the distribution $F$ of the valuations $v$ of the bidders. Since it has a weaker informational requirement, this type of mechanism can be applied in a larger variety of settings.

A mechanism $\Mech$ is a {\bf prior-free $\bm\beta$-approximation} to a benchmark (Def. 7.1. of \cite{MDnA})
 $\Bench$ if for all valuation profiles $v$ its performance is at least a $\beta$ fraction of the benchmark: $\Mech(v) \geq \frac{1}{\beta} \Bench(v)$.

In \cite{DOA} auctions are the mechanisms of interest, in our setting though, we seek to analyze the whole one-shot learner.

\subsection{Differential Privacy}
Differential privacy \citep{CNS}\footnote{The definitions used here are lifted from \cite{DLDP}. For a more in depth exposition of $(\eps, \delta)$-differential privacy see Def. 2.4 of \cite{AFDP} and onwards.} gives us strong guarantees on the distinguishability of the outcomes of a mechanism executed on adjacent datasets.
A randomized mechanism $\Mech \colon \Domain \to \Range$ with domain $\Domain$ and range $\Range$ satisfies {\bf $\bm{(\eps,\delta)}$-differential privacy} if for any two adjacent inputs $d, d'\in \Domain$ and for any subset of outputs $S\subseteq\Range$ it holds that
\begin{equation}\label{eq:diff-priv}
  \Pr[ \Mech (d) \in S] \leq e^{\eps} \Pr[\Mech (d') \in S ]+\delta
\end{equation}

For the auction design context, we take adjacent datasets to be bid profiles which differ in one entry, i.e. it would be contained in one and missing from the other.

A single application of a {\bf gaussian noise mechanism} \begin{equation}
  \label{eq:gmech}
  \Mech(d) \eqdef f(d)+ \NDist(0, S_f^2\cdot \sigma^2)
\end{equation}
to any function $f$, with sensitivity $S_f$ being the amount any single argument can change its output, and $\sigma$ the noise multiplier, satisfies $(\eps, \delta)$-differential privacy if $\delta \geq \frac{4}{5} e^{\frac{-\sigma \eps^2}{2}}$ and $\eps \geq 1$. Since the analysis can be applied post hoc, there are infinite $(\eps, \delta)$ pairs s.t. this is fulfilled.

\subsection{Deep Learning with Differential Privacy}
To introduce differential privacy to deep learning, we can use a differentially private version of SGD, which has two extra steps between the computation of per example gradients $g(x_i)$ and descent.
The gradients get clipped to the gradient norm bound $C$ (which bounds sensitivity $S_f$): 
\begin{equation}
  \bar{g}(x_i) = g(x_i) / \max\Big(1, \frac{\| g(x_i)\|_2}{C}\Big)
\end{equation} Then noise is added to the gradient of each iteration, with $\sigma$ being the noise multiplier, $L$ being the group size of the sample from the whole dataset:
\begin{equation}
  \tilde{g} = \frac{1}{L}\left( \sum_i \bar{g}(x_i) + \mathcal{N}(0, \sigma^2 C^2 \Id)\right)
\end{equation}

This way, training of neural networks can be modelled as a composition of single applications of the gaussian mechanism \eqr{eq:gmech}. For details see Section 3.1. of \cite{DLDP}.

The bookkeeping of the privacy budget resulting from the repeated application of gaussian mechanisms is handled with a moments accountant, which is introduced in \cite{PINQ}, Section 3.2. of \cite{DLDP} describes its application to deep learning.

\subsection{Mechanism Design via Differential Privacy}
\cite{MDvDP} introduce differential privacy as a solution concept for mechanism design problems.

 For any mechanism $\Mech$, truthful reporting is an {\bf $\bm\eps$-approximately dominant strategy} (Definition 10.2 of \cite{AFDP}) for player $i$ if for every pair of types, $t_i, t_i' \in T$, $t_i$ being the private information player $i$ holds, and for every vector of types $t_{-i}$ from the other players, and utility function $u$ the following holds (in the auction setting, types are valuations):
\begin{equation}\label{eq:approx-truth}
  u(t_i, \Mech(t_i, t_{-i})) \geq u(t_i, \Mech(t_i', t_{-i})) - \eps
\end{equation}

Using \eqr{eq:diff-priv}, at the expected utility for any $\bm(\eps, 0)$-differentially private mechanism $\Mech$ and any non-negative function $g$ of its range, with $d, d' \subset \Domain$ differing only in one data point
\begin{equation}
  \E[g(\Mech(d))] \leq e^{\eps} \E[g(\Mech(d'))]
\end{equation}
we can derive that $(\eps, 0)$-differentially private mechanisms being {\bf $\bm{(\eps, 0)}$-approximately dominant strategy truthful} (Section 2.1 of \cite{MDvDP}), for $\eps \leq 1$ and utilities bounded in $[0, 1]$ (Full proof: Lecture 1, Claim 6, \cite{DPMDL}).

This means, given the above constraints, when using an $(\eps, 0)$-differentially private mechanism, no user can cause a change of more than $\eps$ in their utility.

\section{Our approach}
Here we will explain how to combine the above techniques in practice. We believe our approach to lead to an approximately truthful prior-free one-shot learner, giving agents only bounded incentive to misreport without depending on knowledge of valuation distributions. Thus every bid should be an $\eps$-dominant strategy. These results are not yet conclusive, but preliminary experiments yielded promising results.

\subsection{Prior-Free One-Shot Learning}\label{sec:one-shot}
To work towards learning prior-free approximations of optimal auctions, we attempt to make the RegretNet approach approximately truthful by making the regret computation differentially private.

As we do not assume the existence of any valuation distribution, we can only elicit bids and proceed to do one-shot learning on them, to produce one auction per set of bids. We view the whole learner $L$ as the mechanism to be analyzed, since we learn an auction $A$ from a bid profile $b$ and then apply it to the same bid profile to receive the outcome $o$: $L(b) = A$, then $A(b) = o$. From the perspective of the learner, the bids are treated as valuations.

Concretely, we make two modifications to RegretNet Training (Alg. (1) from \cite{DOA}):
\begin{enumerate}
\item We take one set of reports and use it in every iteration of training. (By turning a bid profile into a $\delta$-distribution and repeatedly sampling from it.)
\item We make the computation of the parameters $w$ for the regret gradient differentially private, by using a differentially private optimizer (Alg. (1) from \cite{DLDP}).
\end{enumerate}
If we would only do the first, we would always learn an auction that overfits the reports, meaning misreports could have a large influence. By introducing differential privacy, overfitting is reduced, resulting in bounded influence of each bidders reports.
{\it Note:} We use uniform distributions $\UDist$ to generate misreports $v_i'$. Since we only use the samples to test a set of auction parameters $w$, this is permissible without assuming valuation distributions.

\begin{algorithm}[H]
  \label{alg:one-shot}
  \KwIn{one bid profile $b$}
  \KwOut{one auction $A = (g^w, p^w)$ with $g^w$ and $p^w$ being neural nets parametrized by $w$}
  \KwParam{auction learning rate $\eta>0$, misreport learning rate $\gamma>0$, Lagrange update rates $\forall t, \rho_t>0$, noise scale $\sigma$, gradient norm bound $C$, ($\eta, \gamma, \rho_t, \sigma, C \in \R$),
    training steps $T$, misreport computation steps $\Gamma$, Lagrange update frequency $Q$, ($T, \Gamma, Q \in \N$),
    set of bidders $N$}

  \KwInit{auction parameters $w^0 \in \R^d$, Lagrange multipliers $\lambda^0 \in \R^n$}

  \For{$t=0,\dots,T$}{
    {\bf Initialize misreports:}  ${v'}_{i} \sim \UDist, i \in N$\;
    \For{$r=0,\dots,\Gamma$}{
      \ForAll{$i \in N$}{
      ${v'}_{i} \gets {v'}_{i} +\gamma \nabla_{v'_i} u^w_i\big(b_i; \big({v'}_{i}, b_{-i}\big)\big)$\;
      }
    }

    \BlankLine

    \ForAll{$i \in N$}{
      {\bf Compute regret gradient:}\;
      $g^t_{i} = \nabla_w\big[u^{w}_i\big(b_i; \big({v'}_{i}, b_{-i}\big)\big) -
      u^{w}_i(b_i; b) \big] \,\Big\vert_{w=w^t}$ \;

      \BlankLine

      \begin{tcolorbox}[colframe=white, colback=black!20, arc=0mm, enlarge left by=-3mm, left=2mm]
        {\bf Clip regret gradient:} \tcc*{Differential Privacy}
        $\bar{g}^t_{i} \gets g^t_{i} / \max\Big(1, \frac{\| g^t_{i}\|_2}{C}\Big)$    \;
        \BlankLine

        {\bf Add gaussian noise:} \;
        $\tilde{g}^t_{i} \gets \bar{g}^t_{i} + \mathcal{N}(0, \sigma^2 C^2 \Id)$ \;

      \end{tcolorbox}
    }
    \BlankLine
    {\bf Compute Lagrangian gradient using \eqr{eq:C-grad-one-shot} and update $w^t$:} \;
    $w^{t+1} \gets w^t \,-\, \eta\nabla_w\, C_{\rho_t} (w^{t}, \lambda^{t})$ \;
    \BlankLine

    {\bf Update Lagrange multipliers once in $Q$ iterations:} \;
    \eIf {$t$ is a multiple of $Q$}{
      $\lambda^{t+1}_i \leftarrow \lambda_i^{t} + \rho_t\,\widetilde{\mathit{rgt}}_i(w^{t+1}), ~~\forall i \in N$ \;
    }{$\lambda^{t+1} \gets \lambda^t$ \;}
  }
\caption{One-Shot RegretNet Training}
\end{algorithm}
The Lagrangian function (Section 4 of \cite{DOA})\footnote{The only difference to RegretNet is that $L=1$.} is
\begin{align}
  \label{eq:C-grad-one-shot}
  \nabla_w C_\rho(w, \lambda^{t}) = &- \sum_{i\in N} \nabla_w p^w_i(b) + \sum_{i\in N} \lambda^t_{i} g_{i}
  +\rho \sum_{i\in N} \widetilde{\mathit{rgt}}_i(w) g_{i}
\end{align}

\begin{align*}
\text{with } g_{i} ~=~ \nabla_w\Big[ \max_{v'_i \in V_i}\,u^w_i\big(b_i; \big(v'_i, b_{-i}\big)\big) - u^w_i(b_i; b)\Big].
\end{align*}
The above algorithm returns one auction, consisting of an allocation $g^w$ and payment function $p^w$, per set of reports.

\subsection{Regret Computation for Analysis}
To empirically evaluate whether this auction learner satisfies approximate truthfulness, and to measure the worst case approximation of an optimal auction, we need to compute a set of auctions with different sets of reports. Then we can calculate regret and worst case revenue or welfare over them.

Since we need to train one auction for each set of bids, any evaluation will be computationally expensive. We decided that the smallest meaningful test would be to measure the regret of a single misreporting agent for a sample of points from the valuation space. To calculate regret for one agent on a single point of the valuation space, we take a valuation sample, and compute the corresponding auction $A_0$. We then construct a set of reports by enumerating all possible misreports  $\forall v'^j_1 \in V_1$ of the first agent, while keeping the valuation reports $v_{-1}=(v_2, \dots, v_n)$ of the other agents fixed. For each set of reports $(v'^j_1, v_{-1})$ we then compute an auction $A_j$, with $j \in J$ and $J = (1, \dots, |M \times \mathcal{V}|)$. Since we only handle additive valuations here, we model the valuation space per agent as $M \times \mathcal{V}$ with $\mathcal{V} = \{0, 1\}$ being the discrete valuations we permit, to make enumeration feasible.

The regret for the first agent, over all auctions $A_j \in \mathcal{A}$, with $A_j = (g^{w_j}, p^{w_j}), \forall j \in J$, utility $u^w_i(v_i;b) = v_i(g^w_i(b)) - p^w_i(b)$, and $w_0$ the initial parameters, then is
\begin{align}
  \label{eq:agent1-regret}
  \mathit{rgt_1}(w) = \max_{j \in J} ~ u_1^{w_j}(v_1; (v'^j_1, v_{-1})) - u_1^{w_0}(v_1;(v_1, v_{-1}))
\end{align}

\begin{algorithm}[h]
  \label{alg:one-shot-eval}
  \caption{Computing regret for One-Shot RegretNet}

  \KwIn{Valuation space $\Theta = N \times M \times \mathcal{V}$ with $\mathcal{V} = \{0, 1\}$}
  \KwParam{number of valuation samples $S$}
  \For{$s=0, \dots, S$}{
    {\bf Sample one set of valuations $v$ from $\Theta$}\;
    {\bf Compute one auction $A_0$ on the valuations:}\;
    $\algr{alg:one-shot} \gets (v)$\;
    \ForAll{{\bf misreports }$v'^j_1 \in V_1$}{
      {\bf Compute $A_j$ with one misreport $v'^j_1$ and valuations of the other agents $v_{-1}$:}\;
      $\algr{alg:one-shot} \gets (v'_1, v_{-1})$ \;
    }
    {\bf Calculate regret for $agent_1$ using \eqr{eq:agent1-regret}}
  }
\end{algorithm}

\subsection{Qualitative analysis of empirical results}

In the following figures, each experiment is visualized by a single line. One experiment corresponds to the application of the auction learner to one valuation sample from the valuation space. For each sample we enumerate the misreports of one agent and train one auction each to calculate maximal regret as well as mininimal revenue.
\begin{figure}[H]
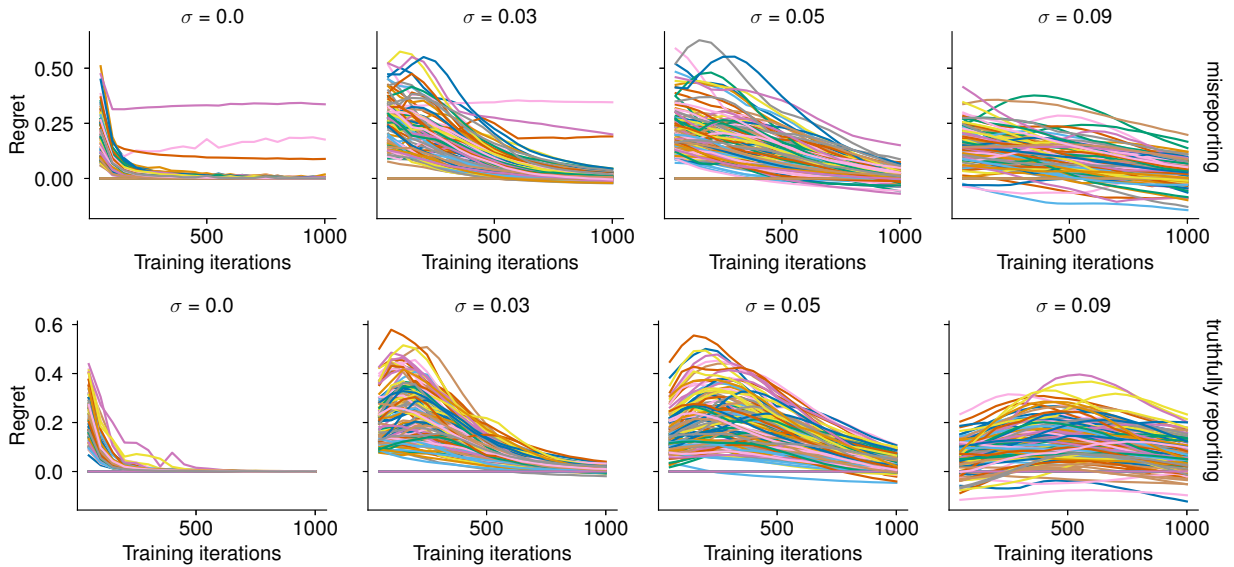

  \centering
  \resizebox{\textwidth}{!}{
    \input{data/5x3/5x3_agent0_sample0.pgf}
  }
  \resizebox{\textwidth}{!}{
    \input{data/5x3/5x3_agent1_sample0.pgf}
  }
  \caption{Max regret of a misreporting and a truthfully reporting bidder (5 bidders, 3 items)}\label{fig:regret}
\end{figure}

\begin{figure}[H]
  \centering
  \resizebox{\textwidth}{!}{
    \input{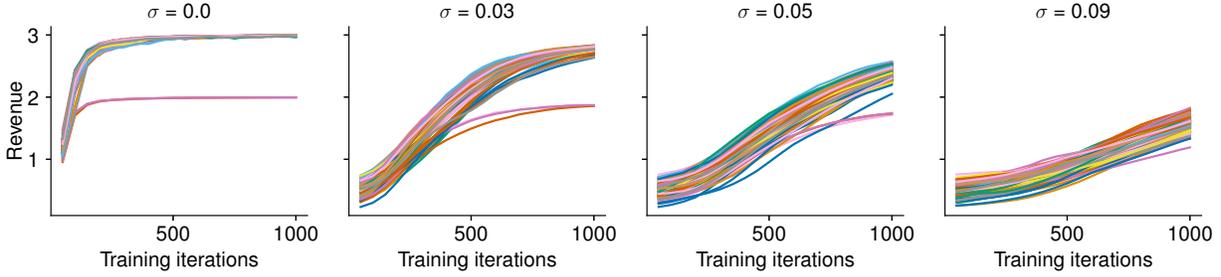}
  }
  \caption{Min revenue (5 bidders, 3 items)}\label{fig:revenue}
\end{figure}

\begin{table}[H]
  \label{tab:priors}
  \centering
  \begin{tabular}{|l|l|r|r|r|r|r|r|}
    \hline
    Learner & Prior & Agents & Items & Regret & Revenue & Iterations & $\sigma$ \\
    \hline
    RegretNet & uniform & 5 & 3 & 0.0022 & 2.10 & 40000 & N/A \\
    one-shot RegretNet & none & 5 & 3 & up to 0.0330 & 1.71 - 2.57 & 1000 & 0.05 \\
    \hline
  \end{tabular}
  \caption{Comparison of prior-free and prior-dependent results}
\end{table}

\begin{table}[H]
  \centering
  \begin{tabular}{|r|r|r|}
    \hline
    $\sigma$ & Regret $agent_1$ & min Revenue overall \\
    \hline
    no DP & 0.337 & 1.993 \\
    0.03  & 0.346 & 1.857 \\
    0.05  & 0.151 & 1.713 \\
    0.09  & 0.198 & 1.191 \\
    \hline
  \end{tabular}
  \caption{Approximation of truthfulness and approximation of outcome optimality depending on $\sigma$}\label{tab:approx}
\end{table}

We can now do a qualitative analysis of the experimental results: In the cases without privacy, misreports that outperform truthful reports exist. When applying differential privacy, we can observe the following stages with increasing noise, dependent on noise multiplier $\sigma$:
\begin{enumerate}
\item More and more misreports stop outperforming truthful reports. Outperforming misreports are visualized by the individual lines outside the main bundles of the misreporting agent in \figr{fig:regret}.
\item At some threshold regret bounds for truthfully reporting and misreporting agents align. This is marked by the bundles of the misreporting and truthfully reporting agent being of roughly the same width from $\sigma = 0.05$ onwards.
\item Regret bounds widen, as can be seen by the bundles of misreporting and truthfully reporting agents increasing in width. Revenue suffers, but its bounds tighten, which is marked by the revenue bundles reducing their width and flattening their slope in \figr{fig:revenue}. This can also be seen in the steady decline of revenue in \tabr{tab:approx}.
\end{enumerate}

This is to be expected, since with differentially private training we exchange privacy for accuracy of the resulting model (see Figure 4, \cite{DLDP}). In the frameworks of \cite{MDvDP, AOMDvDP} privacy controls approximation of truthfulness, accuracy controls outcome quality.

In cases with very small valuation spaces, the technique does not provide reasonable tradeoffs, most likely since noise quickly outweighs available information. Once our pipeline supports finer valuation spaces, we will be able to analyze this in more depth.

All in all, these preliminary results are in line with our hypothesis of being able to train prior-free approximations of optimal mechanisms which are approximately truthful. As this is a work in progress, we have not yet achieved general quantifications for the degrees of approximation for optimality and truthfulness. Producing conclusive evidence will require further investigation.


\section{Further Work}
\subsection{Improving Empirical Analysis}
Since experiments are still comparatively expensive, we don't have good data yet for larger instances, finer valuation spaces or longer training. Application of the technique above to \cite{ALG}, if successful, could make these experiments feasible. An implementation in TensorFlow 2 would allow us to make use of the new SIMD features of TensorFlow Privacy (\cite{FDPSGD}), further reducing training time.

We will also evaluate wether it is possible to use an approximately truthful bid elicitation step for the online algorithm. This might lead to a prior-independent solution, further improving efficiency by being able to train auctions for the approximated distributions, instead of one auction per bid sample.

It might also lead to improved outcomes, if the online learning can be used to turn non-truthful reports into no longer approximately dominant strategies, while preserving approximate dominance for truthful reports (as described on p.3, Lecture 1, \cite{DPMDL}).

\subsection{Approximation Bounds}
Describing bounds for approximation of incentive compatibility, as well as prior-free performance is an open problem for this work and related to the description of generalization bounds for the learning system.

Our current hypothesis is, that to describe meaningful approximation bounds, it is necessary to come up with generalization bounds for iterated, $(\eps, \delta)$-differentially private learning, which are also sensitive to the information capacity of the hypothesis space. To our knowledge, this is still an open problem.

Since the results of the above model depend on the auction setting (different networks are used to model different auctions), as well as on the parameters of the differentially private learning, we can not reuse the generalization bounds from the original model (Sections 2.4 and 3.3 of \cite{DOA}). To get meaningful generalization bounds, one approach would be to extend covering number based techniques, as used in Section 2.4 \cite{DOA} to also account for the learning algorithm that is being used, especially in regard to its privacy, as in \cite{TBDPL}. The resulting technique should bind on the capacity of the hypothesis space, as well as on the sensitivity of the learning.

\subsection{Privacy Aware Agents}
Another interesting topic of further investigation is the influence the technique has on incentive compatibility if the agents are privacy aware \citep{PAMD}.

\section*{Acknowledgements}
Daniel Reusche was funded through the NGI0 PET Fund, a fund established by NLnet with financial support from the European Commission's Next Generation Internet programme, under the aegis of DG Communications Networks, Content and Technology under grant agreement No 825310.

\bibliographystyle{alpha}
\bibliography{bibliography}

\begin{thebibliography}{ACG{\etalchar{+}}16}

\bibitem[ACG{\etalchar{+}}16]{DLDP}
Martín Abadi, Andy Chu, Ian Goodfellow, H.~Brendan McMahan, Ilya Mironov,
  Kunal Talwar, and Li~Zhang.
\newblock Deep learning with differential privacy.
\newblock Oct 2016.
\newblock Proceedings of the 2016 ACM SIGSAC Conference on Computer and
  Communications Security (ACM CCS), pp. 308-318, 2016.

\bibitem[DFN{\etalchar{+}}20]{DOA}
Paul Dütting, Zhe Feng, Harikrishna Narasimhan, David~C. Parkes, and
  Sai~Srivatsa Ravindranath.
\newblock Optimal auctions through deep learning.
\newblock Aug 2020.

\bibitem[DMNS06]{CNS}
Cynthia Dwork, Frank McSherry, Kobbi Nissim, and Adam~D. Smith.
\newblock Calibrating noise to sensitivity in private data analysis.
\newblock In {\em TCC}, 2006.

\bibitem[DR14]{AFDP}
Cynthia Dwork and Aaron Roth.
\newblock The algorithmic foundations of differential privacy.
\newblock {\em Found. Trends Theor. Comput. Sci.}, 9:211--407, 2014.

\bibitem[Har16]{MDnA}
Jason Hartline.
\newblock {\em Manuscript Mechanism Design and Approximation}.
\newblock 2016.

\bibitem[HWT20]{TBDPL}
Fengxiang He, Bohan Wang, and Dacheng Tao.
\newblock Tighter generalization bounds for iterative differentially private
  learning algorithms.
\newblock Aug 2020.

\bibitem[McS10]{PINQ}
Frank McSherry.
\newblock Privacy integrated queries.
\newblock {\em Communications of the {ACM}}, 53(9):89--97, September 2010.

\bibitem[MT07]{MDvDP}
Frank McSherry and Kunal Talwar.
\newblock Mechanism design via differential privacy.
\newblock In {\em 48th Annual IEEE Symposium on Foundations of Computer Science
  (FOCS'07)}, pages 94--103, 2007.

\bibitem[Mye81]{OAD}
Roger Myerson.
\newblock Optimal auction design.
\newblock {\em Mathematics of Operations Research}, 6:58--73, 1981.

\bibitem[NOS12]{PAMD}
Kobbi Nissim, Claudio Orlandi, and Rann Smorodinsky.
\newblock Privacy-aware mechanism design.
\newblock Feb 2012.

\bibitem[NRTV07]{AGT}
Noam Nisan, Tim Roughgarden, Eva Tardos, and Vijay~V. Vazirani.
\newblock {\em Algorithmic Game Theory}.
\newblock Cambridge University Press, 2007.

\bibitem[NST11]{AOMDvDP}
Kobbi Nissim, Rann Smorodinsky, and Moshe Tennenholtz.
\newblock Approximately optimal mechanism design via differential privacy.
\newblock Mar 2011.

\bibitem[RJW20]{ALG}
Jad Rahme, Samy Jelassi, and S.~Matthew Weinberg.
\newblock Auction learning as a two-player game.
\newblock Jun 2020.

\bibitem[Rot14]{DPMDL}
Aaron Roth.
\newblock Lecture notes on differential privacy in game theory and mechanism
  design, Spring 2014.

\bibitem[SVK20]{FDPSGD}
Pranav Subramani, Nicholas Vadivelu, and Gautam Kamath.
\newblock Enabling fast differentially private sgd via just-in-time compilation
  and vectorization.
\newblock Oct 2020.

\end{thebibliography}

\newcommand{\etalchar}[1]{$^{#1}$}

\end{document}